\documentclass[twocolumn,aps,superscriptaddress,showpacs,preprintnumbers,amsmath,amssymb]{revtex4}
\usepackage{graphicx,color}
\usepackage{dcolumn}
\usepackage{colordvi} 
\usepackage{bm}
\usepackage{amssymb}
\begin{document}
\title{Model-independent test of gravity with a network of ground-based gravitational-wave detectors}
\author{Kazuhiro Hayama}
\email{kazuhiro.hayama@ligo.org}
\affiliation{TAMA project, National Astronomical Observatory of
Japan, Mitaka, Tokyo 181-8588, Japan, Albert-Einstein-Institut (Max-Planck-Institut f\"ur Gravitationsphysik), Callinstra\ss e 38, D-30167 Hannover, Germany}
\author{Atsushi Nishizawa}
\email{anishi@yukawa.kyoto-u.ac.jp}
\affiliation{Yukawa Institute for Theoretical Physics, Kyoto University, 
Kyoto 606-8502, Japan}
\date{\today}

\begin{abstract}
The observation of gravitational waves with a global network of interferometric detectors such as advanced LIGO, advanced Virgo, and KAGRA will make it possible to probe into the nature of space\mbox{-}time structure. Besides Einstein's general theory of relativity, there are several theories of gravitation that passed experimental tests so far. The gravitational-wave observation provides a new experimental test of alternative theories of gravity because a gravitational wave may have at most six independent modes of polarization, of which properties and number of modes are dependent on theories of gravity. This paper develops a method to reconstruct an arbitrary number of modes of polarization in time\mbox{-}series data of an advanced detector network. Since the method does not rely on any specific model, it gives model-independent test of alternative theories of gravity. 
\end{abstract}

\maketitle

\section{Introduction}
\label{introduction}

In recent years, direct detection experiments of a gravitational wave (GW) have been well developed and the first generation of a kilometer-scale ground-based laser-interferometric GW detector has accomplished its design sensitivity. Although the first detection of the GW has not been achieved yet, the null detection has yielded scientific results \cite{bib50,2010PhRvD..82j2001A,2010ApJ...713..671A,2009PhRvD..80j2002A,2009PhRvD..80j2001A}. The next-generation interferometers such as advanced LIGO \cite{bib6}, advanced VIRGO \cite{bib7}, and KAGRA \cite{bib8} will be in operation in the coming five years and will bring valuable information about astronomical compact objects.

The direct observation of the GWs will also provide a unique opportunity to test the theory of general relativity (GR), through the propagation speed, waveforms, and polarization modes of GWs. In GR, a GW has two polarization modes (plus and cross modes), while in a general metric theory of gravitation, the GW is allowed to have at most six polarizations \cite{bib1,bib2}. 
In modified gravity theories such as the scalar-tensor theory \cite{bib36,bib37} and $f(R)$ gravity \cite{bib38,bib39}, additional scalar polarizations appear (For more rigorous treatment of the polarizations with the Newman-Penrose formalism, see \cite{bib72,bib73}). 
On the other hand, in bimetric gravity theory \cite{bib40} and massive gravity theory \cite{bib41,bib42}, there appear at most six and five polarization modes, respectively, including scalar and vector modes \cite{bib73,bib43}.
If the additional polarizations are found, it indicates that the theory of gravitation should be extended beyond GR and excludes some theoretical models, depending on which polarization modes are detected. Thus, the observation of the GW polarizations is a powerful tool to probe the extended law of gravity.

Currently, there are few observational constraints on the additional polarization modes of GWs. For the scalar GWs, the observed orbital-period derivative of PSR B1913+16 agrees well with predicted values of GR, conservatively, at a level of $1\,\%$ error \cite{bib44}, indicating that the contribution of scalar GWs to the energy loss is less than $1\,\%$. However, it is important to cross-check the existence of the number of propagating degree of freedom directly by GW detection experiments, since these experiments probe in a weaker regime of gravity than a binary pulsar and at different distance scale. So far some authors have been  the method for separating a mixture of the polarization modes of the GW background and detecting non-Einsteinian polarization modes with pulsar timing array \cite{bib45} and with ground-based laser-interferometric GW detectors \cite{2009PhRvD..79h2002N,bib5}. To decompose and reconstruct the polarization modes, the number of the independent signals of detectors should be more than that of polarization modes. In the above method with GW detectors, the number of polarizations are assumed to be three, since they considered a stochastic background. 

Now we expect that more GW detectors will be in operation in the future, in total five detectors including two advanced LIGOs, advanced VIRGO, KAGRA, and IndIGO \cite{bib9}. Then in principle we can reconstruct five polarization modes (all modes that GW detectors can separately detect) of a transient GW such as a chirp and burst GW. In this paper, we develop a method to separate and reconstruct an arbitrary number of polarization modes using observation data of multiple interferometric GW detectors. The  method does not need any theoretical waveform of the polarization modes of a GW, and therefore can be a model-independent probe of testing various alternative theories of gravity.

The organization of this paper is as follows. In section~\ref{reconstruction} we first describe antenna pattern functions of the tensorial, scalar, and vector polarization modes, and then construct the algorithm to separate and reconstruct the polarization modes based on the coherent network analysis. In section~\ref{implementation}, we describe how the algorithm is implemented and show an example of reconstruction of the polarization modes. In section~\ref{Simulation} we made simulations of the reconstruction of the tensor and scalar polarization modes using simulated data of LIGO Hanford, LIGO Livingston, VIRGO, KAGRA. We devote the last section \ref{summary} to the summary of this paper.

\section{Analytical Method of Reconstruction}
\label{reconstruction}

In this section, we provide the method separately detecting and reconstructing more than three polarization modes of a GW, which often appear in alternative theories of gravity, with a coherent network of ground-based detectors.

\subsection{Polarization modes of a gravitational wave}
In general, a metric gravity theory in four dimensions allows at most six polarization modes of a GW \cite{bib1,bib2}. Let us define a wave orthonormal coordinate that are constructed by a unit vector $\hat{k}$ directed to the propagation direction of a GW and two unit vectors $\hat{e}_{\theta}$ and $\hat{e}_{\phi}$ orthogonal to $\hat{k}$ and each other. With these vectors, the polarization modes are defined as 
\begin{align}
\mathbf{e}^{+} &=\hat{e}_{\theta} \otimes \hat{e}_{\theta} -\hat{e}_{\phi} \otimes \hat{e}_{\phi} \;,\nonumber\\
\mathbf{e}^{\times}&=\hat{e}_{\theta} \otimes \hat{e}_{\phi} +\hat{e}_{\phi} \otimes \hat{e}_{\theta} \;,\nonumber\\
\mathbf{e}^{\circ}&=\hat{e}_{\theta} \otimes \hat{e}_{\theta} +\hat{e}_{\phi} \otimes \hat{e}_{\phi} \;,\nonumber\\
\mathbf{e}^{\ell}&=\sqrt{2}\, \hat{k} \otimes \hat{k}\;,\nonumber\\
\mathbf{e}^{x}&= \hat{e}_{\theta} \otimes \hat{k}+\hat{k} \otimes \hat{e}_{\theta} \;,\nonumber\\
\mathbf{e}^{y}&= \hat{e}_{\phi} \otimes \hat{k}+\hat{k} \otimes \hat{e}_{\phi} \;\nonumber,
\end{align}
where the symbol $\otimes$ denotes a tensor product. The $+$, $\times$, $\circ$, $\ell$, $x$, and $y$ polarization modes are called plus, cross, breathing, longitudinal, vector-x, and vector-y modes, respectively. According to rotation symmetry around the propagation axis of the GW, the $+$ and $\times$ modes are identified with tensor-type (spin-2) GWs, the $x$ and $y$ modes are vector-type (spin-1) GWs, and the $\circ$ and $\ell$ modes are scalar-type (spin-0) GWs. For the more detailed introduction, see \cite{2009PhRvD..79h2002N}. With the sky direction of a GW source $\hat{\Omega}=-\hat{k}$, then a GW with the six polarizations is expressed as
\begin{equation}
h_{ij}(t,\hat{\Omega})= \sum_A h_{A} (t) e_{ij}^{A} (\hat{\Omega})\;,\nonumber
\end{equation} 
where $A=+,\, \times,\, \circ,\, \ell,\, x,\, y$.

\subsection{Antenna pattern functions of polarization modes}
\label{Antenna pattern functions of polarization modes}

Antenna pattern functions of the scalar and vector modes have been derived in \cite{PhysRevD.62.024004,PhysRevD.59.102002,2009PhRvD..79h2002N}. We briefly summarize the results here.

\begin{figure*}
\begin{center}
\includegraphics[width=.49\linewidth]{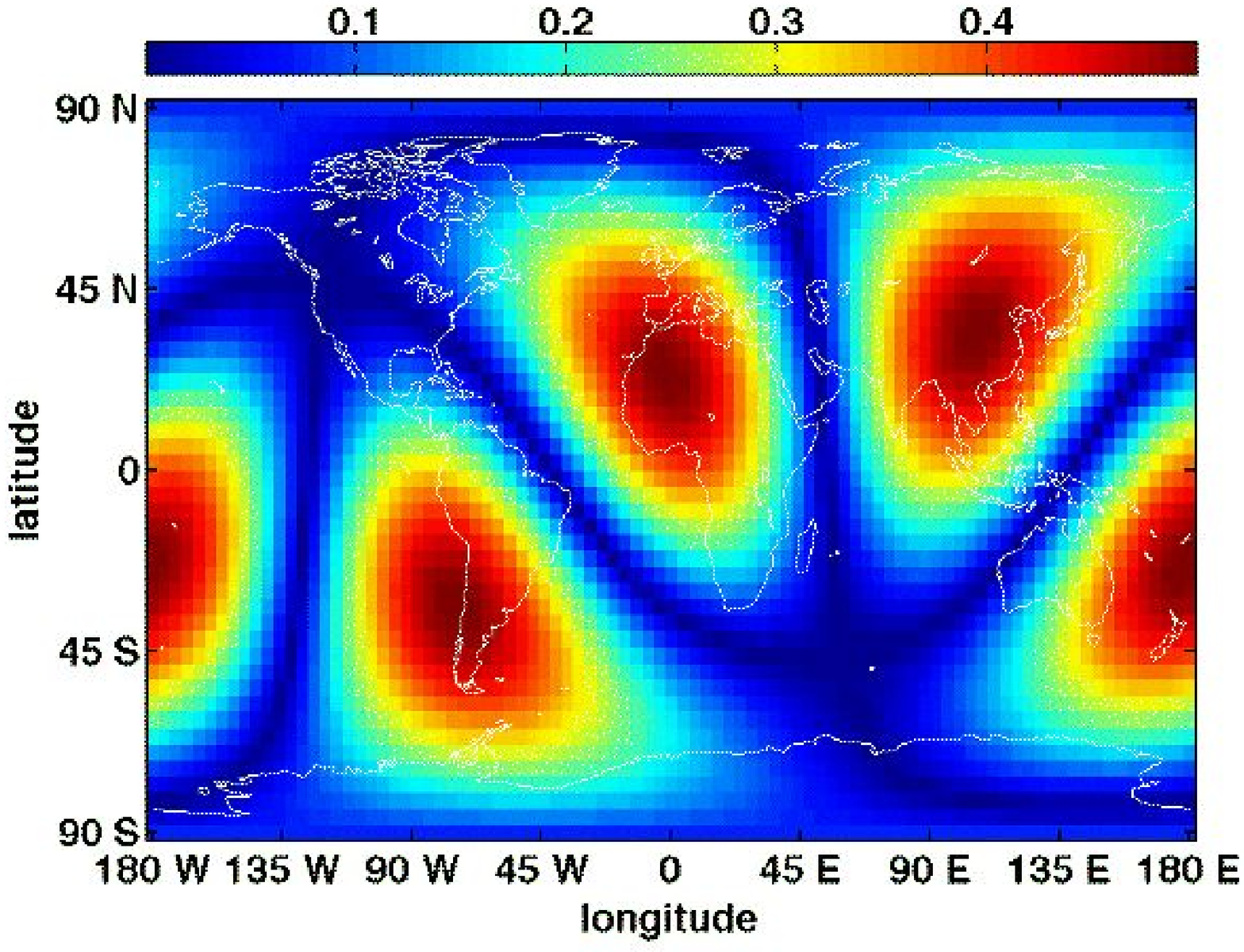}
\includegraphics[width=.49\linewidth]{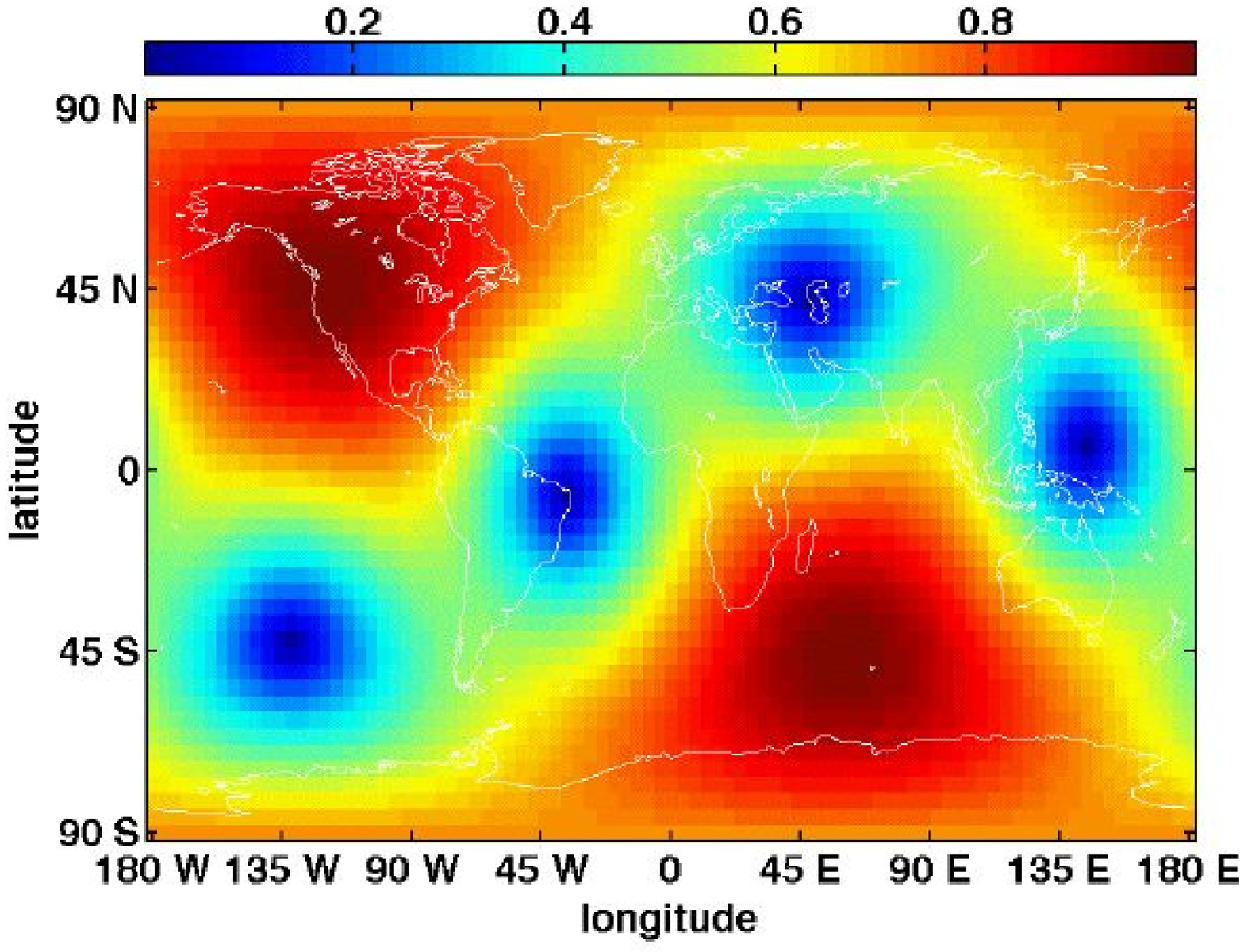}
\includegraphics[width=.49\linewidth]{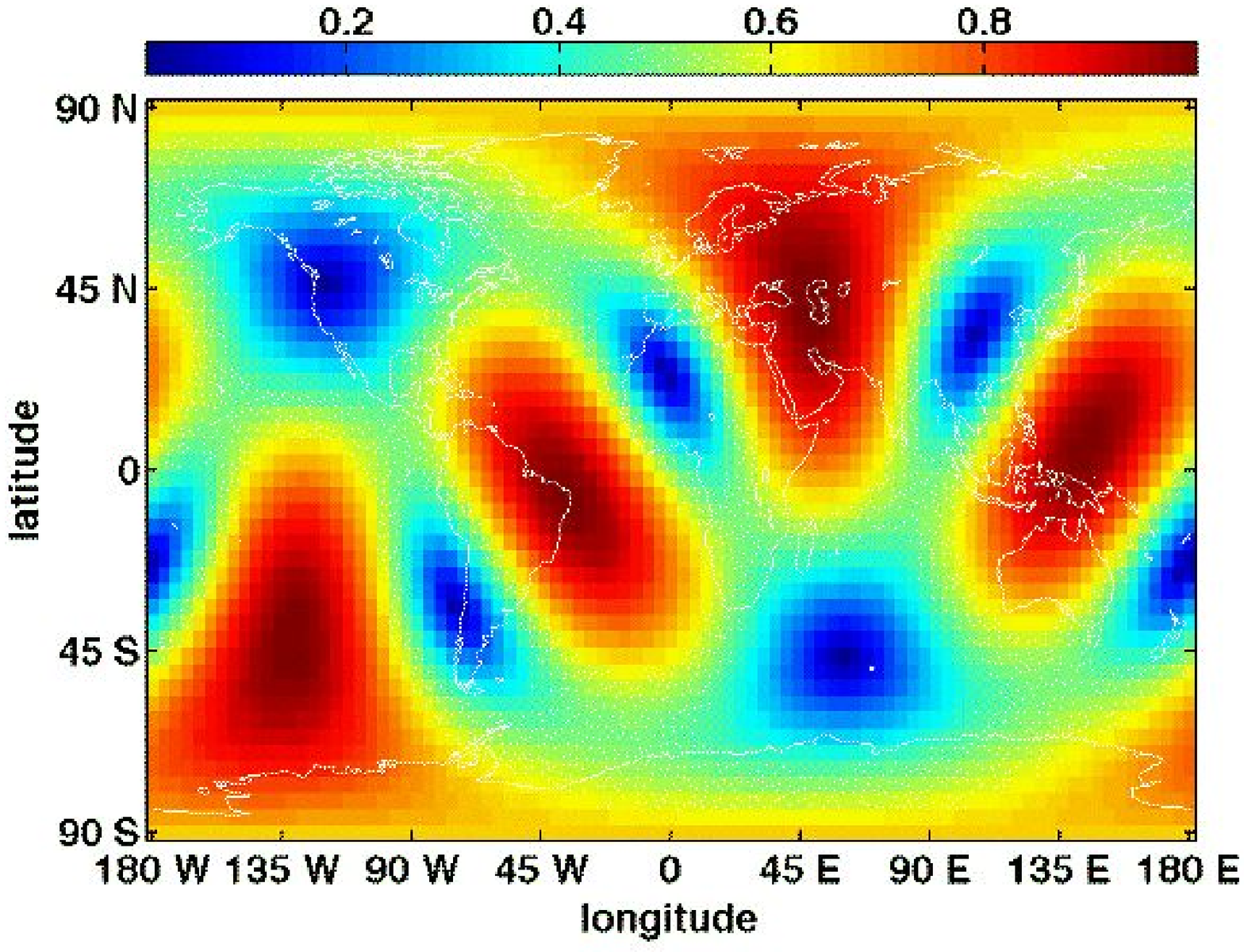}
\caption{The left top plot is an antenna pattern of LIGO interferometric
detector at Hanford to a scalar GW $(F_{\circ}^2)^{1/2}$.  The $X$-axis
is the longitude and the $Y$-axis is the latitude.  The right top plot is an antenna
pattern function of the detector to the tensorial modes
$(F_{+}^2+F_{\times}^2)^{1/2}$. The bottom plot is an antenna pattern function of the detector to the vector modes $(F_{x}^2+F_{y}^2)^{1/2}$
\label{fig:antenna_pattern}}
\end{center}
\end{figure*}

Suppose $m$ interferometric GW detectors are in operation. The GW signal of the $I$\mbox{-}th detector is written as
\begin{equation}
\xi_I(t,\hat{\Omega}) = \sum_A F_I^A (\hat{\Omega}) h_{A}(t) \;,\nonumber
\end{equation}
where $F_I^A (\hat{\Omega})$ is the antenna pattern function of the
$I$\mbox{-}th detector defined as
\begin{equation}
F_I^A (\hat{\Omega}) =e_{ij}^A (\hat{\Omega}) d_I^{ij}\;.\nonumber
\end{equation}
$d_I$ is a detector tensor defined as
\begin{equation}
d_I :=\frac{1}{2} \left[ \hat{u}_I \otimes \hat{u}_I - \hat{v}_I \otimes \hat{v}_I \right],\nonumber
\end{equation}
where $\hat{u}_I, \hat{v}_I$ are unit vectors along with arms of the $I$\mbox{-}th detector. In a spherical coordinate $(\theta_I, \phi_I)$ fixed to the $I$\mbox{-}th detector, since the detector coordinate $(\hat{u},\hat{v},\hat{w})$ and the wave coordinate $(\hat{e}_{\theta},\hat{e}_{\phi},\hat{\Omega})$ are related by
\begin{align}
\hat{e}_{\theta}&=\hat{u}_I \cos \theta_I \cos \phi_I +\hat{v}_I \cos \theta_I \sin \phi_I -\hat{w}_I \sin \theta_I,\nonumber\\
\hat{e}_{\phi}&=-\hat{u}_I \sin \phi_I + \hat{v}_I \cos \phi_I  \;, \nonumber\\
\hat{\Omega} &= \hat{u}_I \sin \theta_I \cos \phi_I +\hat{v}_I \sin \theta_I \sin \phi_I +\hat{w}_I \cos \theta_I,\nonumber
\end{align}
the angular pattern functions for each polarization are 
\begin{align}
F_I^{+}(\hat{\Omega}) &= \frac{1}{2} (1+ \cos ^2 \theta_I) \cos 2\phi_I \;, \nonumber\\
F_I^{\times}(\hat{\Omega}) &= -  \cos \theta_I \sin 2\phi_I \;, \nonumber\\
F_I^{\circ}(\hat{\Omega}) &= -\frac{1}{2} \sin^2 \theta_I \cos 2\phi_I \;, \label{eq1} \\
F_I^{\ell}(\hat{\Omega}) &= \frac{1}{\sqrt{2}} \sin^2 \theta_I \cos 2\phi_I \;, \label{eq2} \\ 
F_I^{x}(\hat{\Omega}) &= -\frac{1}{2} \sin 2\theta_I \cos 2 \phi_I \;, \nonumber\\
F_I^{y}(\hat{\Omega}) &= \sin \theta_I \sin 2\phi_I \;. \nonumber
\end{align}
Note that the wave coordinate has a rotational degree of freedom $\psi$ about the $\hat{\Omega}$ axis. The general formulas keeping $\psi$ arbitrary are provided in \cite{2009PhRvD..79h2002N}. From the expressions, one can see explicit rotational symmetries about $\psi$ for the scalar, vector, and tensor polarization modes. In this paper we take $\psi=0$ without loss of generality. Also we note that the above expressions are valid when the arm length of the detector, $L$, is much smaller than the wavelength of observed GWs, $\lambda_{\rm{g}}$, i.e., $L\ll\lambda_{\rm g}$. This condition is well satisfied for ground-based detectors we consider in this paper. The angular pattern functions for $L > \lambda_{\rm g}$ have been derived in \cite{bib3,bib4} (For pulsar timing, see \cite{bib46,bib47}). Figure \ref{fig:antenna_pattern} shows the scalar, vector, and tensor antenna patterns as a function of the longitude and the latitude.

The angular pattern functions for scalar modes in Eqs.~(\ref{eq1}) and (\ref{eq2}) are degenerated in the long wavelength limit \footnote{In the short wavelength limit, this degeneracy can be broken due to different frequency dependence of the response functions} \cite{bib3,bib4}. These GW signals of the scalar modes cannot be distinguished in GW observation and what we actually detect is the signal combination $\frac{1}{2}F_I^{\circ} \{ h_{\circ}(t) -\sqrt{2}h_{\ell}(t) \}$. From this reason, hereafter we assume, for simplicity, that the longitudinal mode is absent. In addition, we assume that all polarization modes are massless since separate analysis is needed for the mode that propagates with the speed much different from that of light. Therefore we consider in total five massless polarization modes ($+,\, \times,\, \circ,\, x,\, y$) of a GW in the following sections.

\subsection{Coherent network analysis}
\label{Coherent network analysis}

The coherent network analysis is an algorithm to find a transient GW signal, including burst-like signal and in-spiral signal, in the data by combining all available detector data coherently (\cite{Gursel:1989,Klimenko:2005,Mohanty:2006,2006CQGra..23S.673R,2007CQGra..24S.681H,2010NJPh...12e3034S} and therein). The detector output is a combination of  GWs in the polarization modes weighed by the pattern function of each polarization mode. In the coherent network analysis, the sky location of the GW and the waveforms in all polarization modes are reconstructed by inverting the set of the detector responses.

Data $x_I(t)$ from the $I$\mbox{-}th detector is
\begin{equation}
x_I(t)=\xi_I (t)+\eta_I(t)\;,\nonumber
\label{eq:data}
\end{equation}
where $\xi_I (t)$ and $\eta_I(t)$ are the GW signal and noise of the $I$\mbox{-}th detector. The noise is assumed to be Gaussian distributed. The arrival time of a GW at each detector is delayed depending on the geographical locations of the GW detectors. If the relative time delay with respect to a reference time $t_0$ taken at the center of the Earth is defined as $\tau_I(\phi,\theta)$, the arrival time can be redefined as $t=t_0+\tau_I(\phi,\theta)$. 


All the detectors being taken into account, Eq.~(\ref{eq:data}) is written as
\begin{equation}
{\boldsymbol x} = \boldsymbol{F}\boldsymbol{h} + \boldsymbol{\eta},\nonumber
\label{eq:matrixformula}
\end{equation}
where
\begin{eqnarray}
\boldsymbol{x}&=&(\boldsymbol{x}_1,\cdots,\boldsymbol{x}_m)^T \in W[m\times N], \nonumber \\
\boldsymbol{F}&=&\left(
\begin{array}{ccccc} 
\boldsymbol{F}_{1}^{+} & \boldsymbol{F}_{1}^{\times} & \boldsymbol{F}_{1}^{\circ} &\boldsymbol{F}_{1}^{x} & \boldsymbol{F}_{1}^{y} \\
\vdots & \vdots & \vdots & \vdots & \vdots \nonumber \\
 \boldsymbol{F}_{m}^{+} & \boldsymbol{F}_{m}^{\times} & \boldsymbol{F}_{m}^{\circ} &\boldsymbol{F}_{m}^{x} & \boldsymbol{F}_{m}^{y} 
\end{array}
\right) \nonumber \\
&:=& (\boldsymbol{F}_{+},\boldsymbol{F}_{\times},\boldsymbol{F}_{\circ},\boldsymbol{F}_{x},\boldsymbol{F}_{y})\in W[m\times 5], \nonumber \\
\boldsymbol{h}&=&(\boldsymbol{h}_+,\boldsymbol{h}_{\times},\boldsymbol{h}_{\circ},\boldsymbol{h}_{x},\boldsymbol{h}_{y})^T\in W[5\times N], \nonumber \\
\boldsymbol{\eta} &=& (\boldsymbol{\eta}_1,\cdots,\boldsymbol{\eta}_m)^T\in W[m\times N], \nonumber
\end{eqnarray}
and
\begin{eqnarray}
\boldsymbol{x}_I&=&(\hat{x}_I(f_0),\cdots,\hat{x}_I(f_{N-1}))\in W[1\times N]\;, \nonumber \\
\boldsymbol{\eta}_I&=&(\hat{\eta}_I(f_0),\cdots,\hat{\eta}_I(f_{N-1}))\in W[1\times N] \;, \nonumber \\
\boldsymbol{h}_A&=&(\tilde{h}_A(f_0),\cdots,\tilde{h}_A(f_{N-1}))\in W[1\times N]\;, \nonumber 
\end{eqnarray}
$m$ is the number of detectors, $N$ is the number of samplings, and $W[n_1\times n_2]$ represents a type of the matrix with the column $n_1$ and the row $n_2$.
The subscripts run as $I=1,\cdots, m$ and $A=+,\, \times,\, \circ,\, \ell,\, x,\, y$. The components are defined in Fourier space as $\hat{x}_I(f_j):=\tilde{x}_I(f_j)/\sqrt{S_\mathrm{n}^I(f_j)}$, 
and $\hat{\eta}_I(f_j):=\tilde{\eta}_I(f_j)/\sqrt{S_\mathrm{n}^I(f_j)}$, 
$j=0,\cdots,N-1$. $\tilde{x}(f_j)$ is defined as $j$\mbox{-}th component of the Fourier transform of $x(t)$:
\begin{equation}
\tilde{x}(f_j):=\sum_{k=0}^{N-1}x(j\Delta_t)\exp(-2\pi \mathrm{i}k \Delta_t f_j).\nonumber
\end{equation}	
$\Delta_t$ is a sampling period, $\Delta_f:=(N\Delta_t)^{-1}$ is a frequency resolution, and $f_j=j \Delta_f$. $\hat{x}_I(f_j)$, the whitened $\tilde{x}(f_j)$, is obtained by dividing by the power spectrum density $S_\mathrm{n}^I(f_j)$ of $I$\mbox{-}th detector at a frequency $f_j$. There are several purposes to whiten detector data. The sensitivity of an interferometric GW detector is limited in a frequency-dependent way through a diverse noise budget \cite{LIGOSRD}. 
Since multiple data streams have different sensitivities in the Fourier domain, one needs to whiten the data streams so that the noise is isotropically distributed in the space of the detectors. 
The whitening procedure is also important to make the data un-correlated between samples by a whitening filter, and to mitigate the effect of instrumental artifacts in detector data. For instance, noise artifacts that appear in multiple detectors with the same frequency regions can be correlated noise between detectors, which makes the detection efficiency decrease.

We first consider the reconstruction of waveforms in an ideal case without noise. In general, $\boldsymbol{F}$ is not a squared matrix with full rank, hence we introduce the Moore-Penrose pseudo-inverse matrix $\boldsymbol{M}$ as
\begin{equation}
\boldsymbol{M}\boldsymbol{h}=\boldsymbol{F}^T\boldsymbol{x}, \quad \mathrm{where}  \quad\boldsymbol{M}:=\boldsymbol{F}^T\boldsymbol{F}.\nonumber
\end{equation}
If the detectors are not all co-aligned, $\boldsymbol{M}$ is an invertible $5\times 5$ matrix. Multiplying the equation by the inverse of $\boldsymbol{M}$, we get
\begin{equation}
	\boldsymbol{h}=\boldsymbol{F}^{\dagger}\boldsymbol{x}, \quad\mathrm{where}\quad \boldsymbol{F}^\dagger:=\boldsymbol{M}^{-1}\boldsymbol{F}^T.\nonumber
\end{equation}
The inverse matrix $\boldsymbol{M}^{-1}$ can be found from the formula
\begin{equation}
	\boldsymbol{M}^{-1}=\frac{1}{\mathrm{det}(\boldsymbol{M})} \mathrm{adj}(\boldsymbol{M}),\nonumber
\end{equation}
where $\mathrm{adj}(\boldsymbol{M})$ is the adjoint matrix and $\mathrm{det}(\boldsymbol{M})$ is the determinant of $\boldsymbol{M}$.
If $(i,j)$-th cofactor is defined as $\boldsymbol{C}_{AA'}=[\mathrm{adj}(\boldsymbol{M})]_{AA'}$, we finally obtain
\begin{eqnarray}
\boldsymbol{h}_A&=&\boldsymbol{H}_A\cdot\boldsymbol{x}\;,
\label{eq:hrecon} \\
\boldsymbol{H}_A &=& \frac{1}{\mathrm{det}(\boldsymbol{M})} \sum_{A'} \boldsymbol{C}_{AA'} \cdot \boldsymbol{F}_{A'} \label{eq3}
\end{eqnarray}
As discussed in \cite{bib5}, the factor $\mathrm{det}(\boldsymbol{M})$ of this formula plays an important role in separating the polarization modes. If there is degeneracy in the antenna pattern functions of a detector network and $\mathrm{det}(\boldsymbol{M})=0$, we cannot reconstruct the polarization modes at all. Fortunately, this is not the case for current ground-based detectors.



We use a maximum likelihood method to estimate $\boldsymbol{h}$ from the data. The maximum likelihood method maximizes
\begin{eqnarray}
L[\boldsymbol{h}]:&=&-\parallel\boldsymbol{x}-\boldsymbol{F}\boldsymbol{h}\parallel^2,\\
&=&-\parallel\boldsymbol{x}-\boldsymbol{F}\boldsymbol{F}^{\dagger}\boldsymbol{x}\parallel^2,\label{eq:ML}
\end{eqnarray}
where $\parallel\cdot\parallel$ is defined by
\begin{equation}
\parallel\boldsymbol{x}\parallel:=\left[ \sum_{I=1}^m\sum_{j=0}^{N-1} | \hat{x}_I (j\mathnormal{\Delta}_f)|^2 \mathnormal{\Delta}_f \right]^{1/2}.\nonumber
\end{equation}
Introducing $\boldsymbol{Q}:=\boldsymbol{I}-\boldsymbol{F}\boldsymbol{F}^{\dagger}$ and using $\boldsymbol{Q}$ in Eq.~(\ref{eq:ML}), we obtain
 \begin{equation}
L[\boldsymbol{h}]=-\parallel \boldsymbol{Q}\boldsymbol{x}\parallel^2,\\
\label{eq:Qlikelihood}
\end{equation}
We note $\boldsymbol{Q}\boldsymbol{F}=0$, which means that $\boldsymbol{Q}$ projects onto the null space of $\boldsymbol{F}\boldsymbol{F}^T$. From Eq.~(\ref{eq:Qlikelihood}), one can see $L[\boldsymbol{h}]$ is equivalent to the null stream energy~\cite{2006PhRvD..74h2005C,2012PhRvD..86b2004C}. Suppose the true source location is $\hat{\Omega}_s:=(\phi_s,\theta_s)$,
\begin{align}
	L[\boldsymbol{h}]&=-\parallel \boldsymbol{Q}(\boldsymbol{F}(\hat{\Omega}_s)\boldsymbol{h}+\boldsymbol{\eta})\parallel^2\nonumber\\
	&=-\parallel \boldsymbol{Q}\boldsymbol{F}(\hat{\Omega}_s)\boldsymbol{h}\parallel^2-2 {\rm{Re}} \left[ \sum_j (\boldsymbol{\eta}^{\dagger}\boldsymbol{Q}\boldsymbol{F}(\hat{\Omega}_s)\boldsymbol{h})_j \mathnormal{\Delta}_f \right] \nonumber \\
&\quad -\sum_j (\boldsymbol{\eta}^{\dagger}\boldsymbol{Q}\boldsymbol{\eta})_j \mathnormal{\Delta}_f.  \label{eq:Qlikelihood2} 
\end{align}
We used the relation $\boldsymbol{Q}^2=\boldsymbol{Q}$ and denoted frequency components as $(\boldsymbol{\eta}^{\dagger}\boldsymbol{Q}\boldsymbol{\eta})_j = [\boldsymbol{\eta}^{\dagger}\boldsymbol{Q}\boldsymbol{\eta}](f_j)$. As Ref.~\cite{2006CQGra..23S.673R} shows, if we take the expectation value of the likelihood function in a stationary noise case, the first term in Eq.~(\ref{eq:Qlikelihood2}) is the same as the current value $\parallel \boldsymbol{Q}\boldsymbol{F}(\hat{\Omega}_s)\boldsymbol{h}\parallel^2$ and the second term vanishes. The third term at $j$\mbox{-}th frequency bin is 
\begin{eqnarray}
\langle (\boldsymbol{\eta}^{\dagger}\boldsymbol{Q}\boldsymbol{\eta})_j\rangle&=&\sum^m_{I,J} Q^j_{IJ} \langle (\hat{\eta}_I^j)^{\ast} \hat{\eta}_J^j \rangle \nonumber \\
&=& \sigma_j^2 \sum^m_{I,J} \delta_{IJ} Q^j_{IJ}  \nonumber \\
&=&\sigma_j^2\, \mathrm{tr}(\boldsymbol{Q}^j),\nonumber 
\end{eqnarray}
where $\delta_{IJ}$ is Kronecker delta, and $\langle \cdot \rangle$ is the ensamble average. Since the noise in the detectors is whitened, the noise can be regarded as the Gaussian white noise with its variance $\sigma^2 := \sum_j \sigma_j^2 \Delta_f$, where $\sigma_j^2$ is defined by $\langle (\hat{\eta}_I^j)^{\ast} \hat{\eta}_J^j \rangle = \delta_{IJ} \sigma_j^2$. We note $\sigma_j^2$ has the same value for all $j$\mbox{-}th frequency bin since the noise $\hat{\eta}$ is whitened. Transforming $\boldsymbol{Q}^j$ to the diagonal form, we find $\mathrm{tr}(\boldsymbol{Q}^j)=m-n$ and finally obtain	
\begin{equation}	
	\langle L[\boldsymbol{h}]\rangle =-\parallel \boldsymbol{Q}\boldsymbol{F}(\hat{\Omega}_s)\boldsymbol{h}\parallel^2- (m-n) \sigma^2,
\label{eq:Qlikelihood3}
\end{equation}
where $n$ is the total number of polarization modes. Since $\boldsymbol{Q}\boldsymbol{F}(\hat{\Omega}_s)=\boldsymbol{0}$ at the true location, $L[\boldsymbol{h}]$ become maximum.

We emphasize that the algorithm does not specify any alternative theories of gravity and can be applied to any number of polarization modes if it is less than the number of detectors. Therefore, this approach is a model-independent probe alternative theories of gravity. In addition, we here considered the transient GW signals, but in principle the method works in case of a continuous GW signal by considering the signal modulation due to the earth rotation and time dependence of the antenna patern functions.

\section{Application to three polarization modes}

\subsection{Scalar GW in Brans-Dicke theory}
\begin{figure}
\begin{center}
\includegraphics[width=1\linewidth]{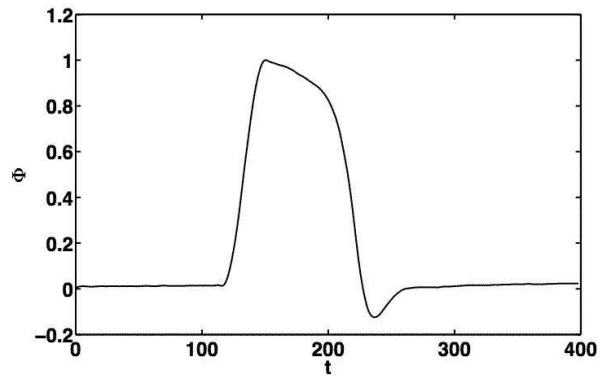}
\caption{
A scalar gravitational waveform simulated by Shibata et al.\cite{shibata:1994} with the mass of the progenitor $10\mathrm{M_{\odot}}$, $\omega_{\mathrm{BD}}=40000$, the distance from the earth $10\mathrm{Mpc}$. The unit of the time $t$ is $4.93\times10^{-5}(M/10M_{\odot})$. $\Phi$ is the normalized strain amplitude.
\label{fig:scalarGW}
}
\end{center}
\end{figure}
There are several simulations on the spherically symmetric core collapse in the Brans-Dicke theory \cite{bib36}, which is a class of scalar-tensor theory. The scalar gravitational waveform is consistent with each simulation~\cite{shibata:1994,PhysRevD.56.785,1997PhRvD..55.2024H,1998PhRvD..57.4802H,1998PhRvD..57.4789N,2000ApJ...533..392N}. Therefore we use a waveform simulated by Shibata et al.~\cite{shibata:1994}. In their simulation, they assumed the Brans-Dicke theory with the scalar field coupling to gravity, $\omega_{\mathrm{BD}}=50$ and $500$, and calculated scalar gravitational waveforms. They found the scalar fields are linearly scaled with $\omega_{\mathrm{BD}}$, and the scalar gravitational waveforms with different $\omega_{\mathrm{BD}}$ are almost the same by scaling $\omega_{\mathrm{BD}}$. We therefore extrapolate the waveforms to larger values of $\omega_{\mathrm{BD}}$.
The simulated scalar GW signal $h$ is 
\begin{equation}
h(t) = 1.25\times 10^{-21}\left(\frac{M}{10M_{\odot}}\right)\left(\frac{10\mathrm{kpc}}{R}\right)\left(\frac{40000}{\omega_{\mathrm{BD}}}\right)\Phi(t)\nonumber 
\end{equation}
where $M$ is the mass of a progenitor, $R$ is the distance from the earth. The shape of the waveform $\Phi(t)$ is in Figure~\ref{fig:scalarGW}. Since the unit of the time $t$ in Figure~\ref{fig:scalarGW} is $4.93\times10^{-5}(M/10M_{\odot})$, the duration of the signal is linearly dependent of a mass of a progenitor. 

%
\subsection{Sensitivity to the scalar mode}

\begin{figure}
\begin{center}
\includegraphics[width=0.9\linewidth]{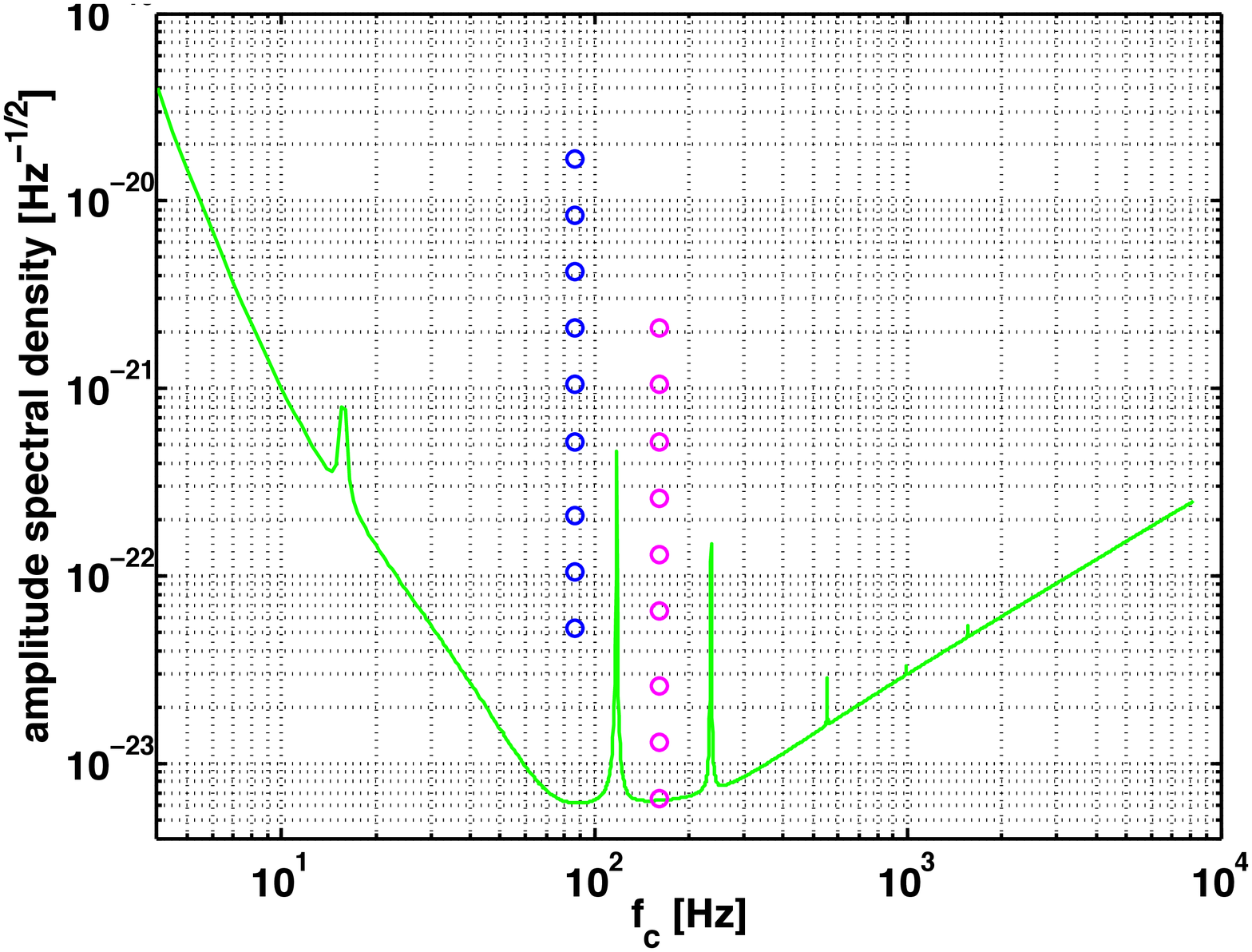}
\includegraphics[width=0.9\linewidth]{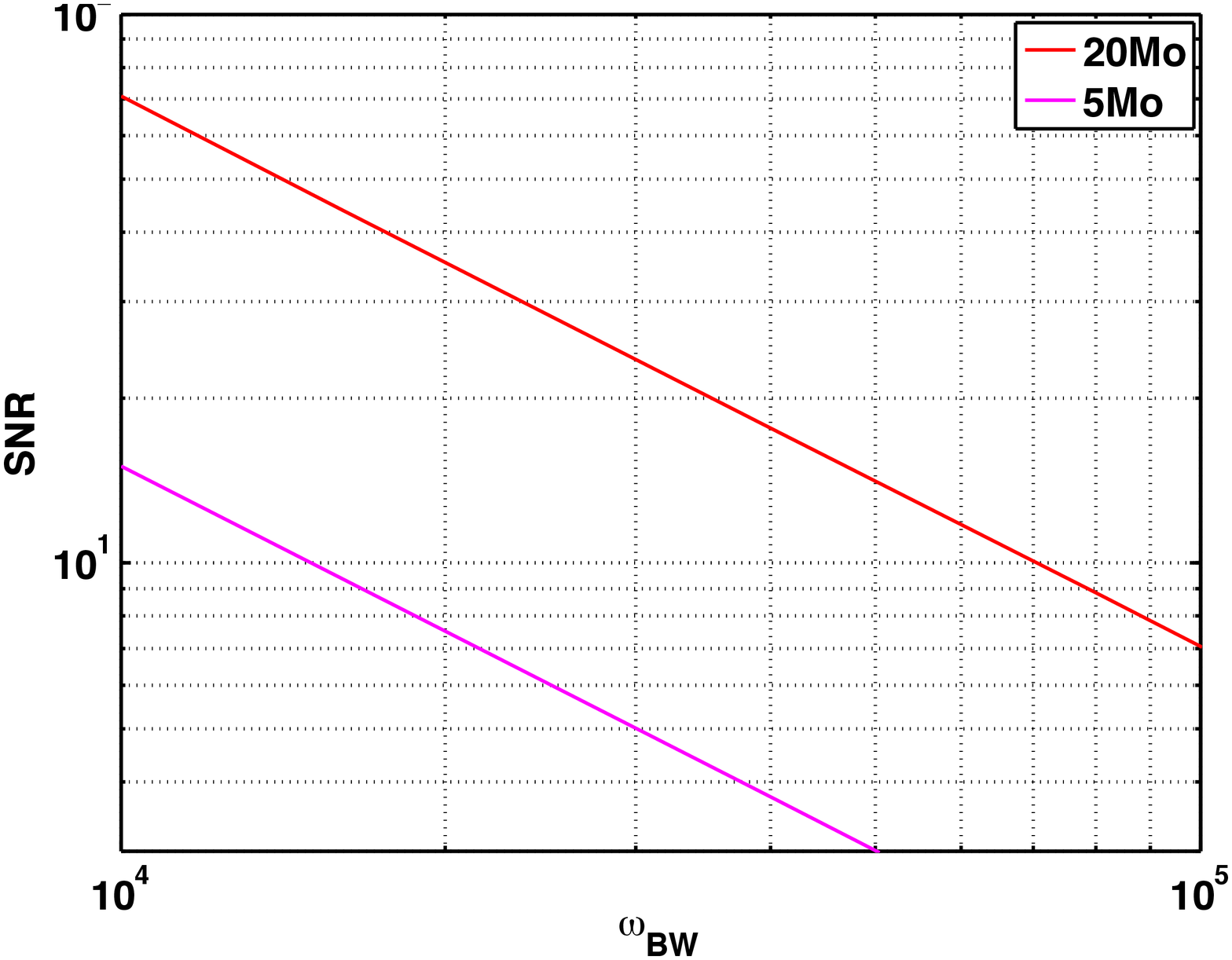}
\caption{The green line is the KAGRA sensitivity curve. Blue open circles are $h_{\mathrm{rss}}$-$f_c$ of spherically symmetric core collapse supernovae at the distance $10\mathrm{kpc}$ with the progenitor mass $20\mathrm{M_{\odot}}$ and $\omega_{\mathrm{BD}}=500,1000,2000,4000,8000,16000,40000,80000,160000$. $h_{\mathrm{rss}}$, $f_c$ are defined in the text. Magenta open circles are for the progenitor mass $5\mathrm{M_{\odot}}$. The bottom plot is $\omega_{\mathrm{BD}}$-$\mathrm{SNR}$. $\mathrm{SNR}$ is defined in the text. The red line is the one with the mass of the progenitor $20\mathrm{M_{\odot}}$, the magenta line $5\mathrm{M_{\odot}}$.
\label{fig:sensitivity2scalar}
}
\end{center}
\end{figure}
Sensitivity of the designed Japanese interferometric GW detector KAGRA~\cite{KAGRA_curve} to a scalar GW simulated by Shibata et al.~\cite{shibata:1994} is estimated. We should note that more realistic estimation of the detectability needs more realistic simulation. However, we use Shibata's result for the purpose of demonstration. The top plot in Fig.~\ref{fig:sensitivity2scalar} shows $h_{\mathrm{rss}}$-$f_c$ of GWs from spherically symmetric core collapse supernovae located at $10$kpc from the earth with $\omega_{\mathrm{BD}}=500,1000,2000,4000,8000,16000,40000,80000,160000$ from the top. $h_{\mathrm{rss}}$ is the root sum square of $h$ defined as
\begin{equation}
h_{\mathrm{rss}}:=\left(\int^{\infty}_{-\infty}\mathrm{d}t\sum_Ah_{A}^2(t)\right)^{1/2},\nonumber 
\end{equation}
$A=+,\times,\circ$, and $f_c$ is the characteristic frequency~\cite{Thorne:1987} 
\begin{eqnarray}
f_c:=\left(\int^{\infty}_0\frac{\sum_A\tilde{h}_A(f)\tilde{h}_A^*(f)}{S_\mathrm{n}(f)}f\mathrm{d}f\right)\times\nonumber \\
\left(\int^{\infty}_0\frac{\sum_A\tilde{h}_A(f)\tilde{h}_A^*(f)}{S_\mathrm{n}(f)}\mathrm{d}f\right)^{-1},\nonumber 
\end{eqnarray}
where $\tilde{x}(f)$, $S_\mathrm{n}(f)$, $\tilde{h}_A(f)$ are detector output, the noise power spectrum density, the plus, cross, and scalar modes of a GW in Fourier domain. $\tilde{h}_A^*$ expresses its complex conjugate. 
Note that here we are considering GWs from spherically symmetric core collapse supernovae, which does not radiate tensor modes, namely, $h_+(t)=h_{\times}(t)=0$.
$h_{\mathrm{rss}}$ of the supernovae with the progenitor mass of $20\mathrm{M_{\odot}}$ at the distance $10\mathrm{kpc}$ are shown with blue open circles. The magenta open circles are the same except the mass of the progenitor is $5\mathrm{M_{\odot}}$. These plots are on the current design sensitivity curve of KAGRA.
The bottom plot in Fig.~\ref{fig:sensitivity2scalar} shows estimated signal-to-noise ratio (SNR) as a function of $\omega_{\mathrm{BD}}$. The SNR is defined as 
\begin{equation}
\mathrm{SNR}:=\left(\int^{\infty}_{-\infty}\frac{\sum_A\tilde{x}(f)\tilde{h}_A^*(f)}{S_\mathrm{n}(f)}\mathrm{d}f\right)^{1/2}.\nonumber 
\label{eq:def_snr}
\end{equation}
We note that the SNR defined here is that of the matched filter method~\cite{1992PhRvD..46.5236F}, which is the optimal case with respect to the antenna pattern function of the detector. 
The SNR is 
estimated by 
\begin{equation}
\mathrm{SNR}\simeq\frac{h_c}{\sqrt{f_cS_\mathrm{n}(f_c)}},\nonumber 
\end{equation}
where $h_c$ is the characteristic strain amplitude~\cite{Thorne:1987}
\begin{equation}
h_c:=\left(3\int^{\infty}_0\frac{S_\mathrm{n}(f_c)}{S_\mathrm{n}(f)}\sum_A\tilde{h}_A(f)\tilde{h}_A^*(f)f\mathrm{d}f\right)^{1/2} \;.\nonumber 
\end{equation}
 The red and magenta lines are the ones with the mass of the progenitor $20\mathrm{M_{\odot}}$ and $5\mathrm{M_{\odot}}$, respectively. The current upper limit on $\omega_{\mathrm{BD}}$ is $> 40000$ by Cassini's observation in the solar system~\cite{2003Natur.425..374B}. If a scalar GW with $\omega_{\mathrm{BD}}=40000$ is radiated from a spherically symmetric core collapse supernova at $10$~kpc from the earth, KAGRA can detect the signal with SNR of $\sim 18$ for the mass of $20\mathrm{M_{\odot}}$ progenitor, $\sim 4$ for the mass of $5\mathrm{M_{\odot}}$ progenitor. 
In order to set upper limit on $\omega_{\mathrm{BD}}$, one has to know the distance of a source because the distance and $\omega_{\mathrm{BD}}$ influence the amplitude of the GW at a detector in the same way. Therefore the two parameters are degenerated. It is important to combine usual astronomical observations such as electro-magnetic observations and neutrino observations with the GW observation, so called the {\it{multi-messenger observation}}. 
Suppose it finds a supernova occurs within our Galaxy through the multi-messenger observation. The above results indicates one may set stronger upper limit on $\omega_{\mathrm{BD}}$, or if the scalar GW is detected, one can determine $\omega_{\mathrm{BD}}$.

\subsection{Reconstruction of three polarization modes}

In the case the tensor and scalar modes exist, $h_+$, $h_\times$, $h_\circ$ are reconstructed as the linear combination of the antenna patterns of all the three polarizations. 
$\boldsymbol{H}_A$ ($A=+, \times, \circ$) can be explicitly calculated from Eq.~(\ref{eq3}):
\begin{eqnarray}
	\boldsymbol{H_+}&=&\frac{1}{\mathrm{det}(\boldsymbol{M})}[
	(\boldsymbol{F_\times}\times \boldsymbol{F_\circ})\cdot(\boldsymbol{F_\times}\times \boldsymbol{F_\circ})\boldsymbol{F_+}\nonumber\\
	&-&(\boldsymbol{F_\times}\times \boldsymbol{F_\circ})\cdot(\boldsymbol{F_+}\times \boldsymbol{F_\circ})\boldsymbol{F_\times}\nonumber\\
	&+&(\boldsymbol{F_\times}\times \boldsymbol{F_\circ})\cdot(\boldsymbol{F_+}\times \boldsymbol{F_\times}) \boldsymbol{F_\circ}],\label{eq:H_+}\\
	\boldsymbol{H_\times}&=&\frac{1}{\mathrm{det}(\boldsymbol{M})}[
	-(\boldsymbol{F_+}\times \boldsymbol{F_\circ})\cdot(\boldsymbol{F_\times}\times \boldsymbol{F_\circ})\boldsymbol{F_+}\nonumber\\
	&+&(\boldsymbol{F_+}\times \boldsymbol{F_\circ})\cdot(\boldsymbol{F_+}\times \boldsymbol{F_\circ})\boldsymbol{F_\times}\nonumber\\
	&-&(\boldsymbol{F_+}\times \boldsymbol{F_\circ})\cdot(\boldsymbol{F_+}\times \boldsymbol{F_\times}) \boldsymbol{F_\circ}],\label{eq:H_x}\\
	\boldsymbol{H_\circ}&=&\frac{1}{\mathrm{det}(\boldsymbol{M})}[
	(\boldsymbol{F_+}\times \boldsymbol{F_\times})\cdot(\boldsymbol{F_\times}\times \boldsymbol{F_\circ})\boldsymbol{F_+}\nonumber\\
	&-&(\boldsymbol{F_+}\times \boldsymbol{F_\times})\cdot(\boldsymbol{F_+}\times \boldsymbol{F_\circ})\boldsymbol{F_\times}\nonumber\\
	&+&(\boldsymbol{F_+}\times \boldsymbol{F_\times})\cdot(\boldsymbol{F_+}\times \boldsymbol{F_\times})\boldsymbol{F_\circ}]\label{eq:H_o}.
\end{eqnarray}
The symbols $\times$ and $\cdot$ denote the outer product and the inner product, respectively. We note if $\mathrm{det}(\boldsymbol{M})=0$, which happens when detectors are co-aligned, the reconstruction of any polarization mode fails. However, this is not the case for the current ground-based detector network.

\section{Simulation of reconstruction of polarization modes}
\label{Simulation}

\subsection{Implementation}
\label{implementation}

The pipeline consists of two parts:
\begin{itemize}
\item Data conditioning
\item Event trigger generation
\item Reconstruction of the poralizations
\end{itemize}
We use the linear prediction error filter for whitening the data, assuming data is a stationary stochastic process and can be expressed by an autoregressive model with order of $P$ where $P$ is an integer number. We assume the predicted data stream at the time $s{\Delta}_t$, $x'(s{\Delta}_t)$ can be written as 
\begin{equation}
x'(s{\Delta}_t)=\sum^P_{p=1}c(p)x((s-p){\Delta}_t),\nonumber 
\end{equation}
where $s$ is an integer number and ${\Delta}_t$ is a sampling period.
Then we try to find parameters $c(p)$ which minimize the mean squared prediction error $E_e^2:= \frac{1}{N}\sum_{s=0}^{N-1}x_\mathrm{w}(s{\Delta}_t)^2$, where $N$ is the number of samples and 
\begin{eqnarray}
x_\mathrm{w}(s{\Delta}_t)&:=&x(s{\Delta}_t)-x'(s{\Delta}_t)\nonumber 
\end{eqnarray}
is a whitened data of $x(s\Delta_t)$. 
The coefficients $c(p)$ are obtained in the lease squares sense by requiring 
\begin{equation}
	\frac{\partial E_e^2}{c(p)}=0, \;\; 0\leq p\leq N-1.\label{eq:lsq}
\end{equation}
This equation results in the Yule-Walker equations \cite{Makhoul_1975}. 
The whitening filter has a group delay of the phase especially near narrowband spectral feature, and the timing error between the whitened data of different detectors is introduced. To cancel the timing error, the linear predictor filter is first applied causally and then anti-causally~\cite{Chatterji_2004}.
The filter coefficients $c(p)$ should be estimated using a stationary data segment. For this purpose, we use a few seconds data segment that does not have non-stationary noise in the implementation. The data segment will not be used for search for GWs. We then construct the finite impulse response (FIR) filter 
 and pass the data through the FIR filter to whiten it.

Set of the whitened data $\boldsymbol{x_\mathrm{w}}$ of the multiple detectors with equal length is passed to the part of the event trigger generation. We use the coherent network analysis pipeline described in \ref{Coherent network analysis}. The value of the likelihood of the multiple detector data defined in Eq.~(\ref{eq:ML}) is calculated by changing over the possible sky locations $\hat{\Omega}=(\theta,\phi)$, and the maximum of the likelihoods is chosen. If the maximum likelihood value is above a given threshold, the chosen event candidate is recorded in a detection list. Since the likelihood values are obtained as a function of $\theta$ and $\phi$, this two-dimensional output, $\boldsymbol{S}(\theta,\phi)$, is called {\it{skymap}}. 

Finally reconstructed $h_{+}$, $h_{\times}$ and $h_{\circ}$ are evaluated the maximum point of the skymap using Eqs.~(\ref{eq:hrecon}), (\ref{eq3}), (\ref{eq:H_+}), (\ref{eq:H_x}), and (\ref{eq:H_o}). The reconstructed $h_{+}$, $h_{\times}$, and $h_{\circ}$ have durations less than or equal to the data length. We note that the implementation is based on the paper~\cite{2007CQGra..24S.681H}.

\subsection{Reconstruction of the polarization modes without tensor mode signals}

\begin{figure}
\begin{center}
\includegraphics[width=1\linewidth]{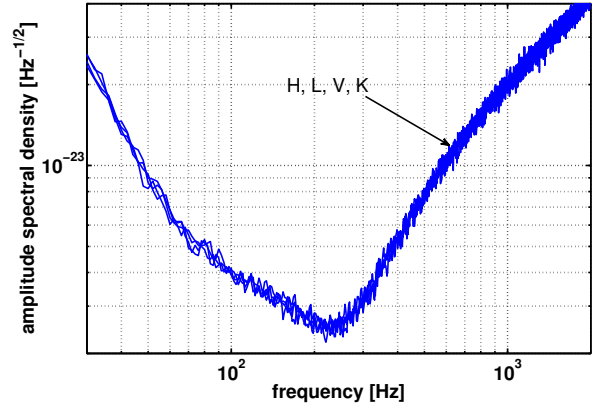}
\caption{The sensitivity curves. The $x$-axis is frequency and the $y$-axis is amplitude spectrum density. Detectors considered are H, L, V, K with the same sensitivity which is similar to the design sensitivity of advanced LIGO.}
\label{fig:detector_sensitivity}
\end{center}
\end{figure}
We perform Monte Carlo simulations to reconstruct the scalar polarization of a GW. The network consisted of the $4$~km LIGO Hanford (H), LIGO Livingston (L), VIRGO (V), and KAGRA (K) interferometers~\cite{bib6,bib7,bib8}. For the detector noise amplitude spectral densities, we use the design sensitivity curves similar to the advanced LIGO detectors as given in~\cite{2009LRR....12....2S} (see Fig.~\ref{fig:detector_sensitivity}) and keep the locations and orientations the same as the
real detectors. The stationary noise is generated $50$~seconds by first using $4$ independent realizations of Gaussian white noise and then passing them through FIR filters having transfer functions that {\em approximately} match the design curves.
The generated data is sampled at $16384$~Hz and then passed through the data conditioning pipeline. Besides downsampling the data to $2048$~Hz by applying the same anti-aliasing filter to all data streams, the data conditioning pipeline applies time domain whitening filters that are trained on the first five seconds of data in which any injected signal is not included. In this simulation we put cut\mbox{-}off frequencies at $60$~Hz and $400$~Hz so that lower and higher frequencies are filtered.

We assume that the GW source is a spherically symmetric core collapse supernovae. The injected signals corresponds to a single source located at the right ascension
(RA) of $15~{\rm hours}$ and the declination (DEC) of $-60~{\rm degrees}$. The mass of the progenitor is $10\mathrm{M_{\odot}}$, $\omega_{\mathrm{BD}}=40000$, the distance from the earth is $10\mathrm{kpc}$. we here assume that $h_\times(t)$ and $h_{\mathrm{+}}(t)$ are absent and that the simulated gravitational waveform is $h_{\circ}$. In the next subsection, we will perform the same simulation in the presence of $h_\times(t)$ and $h_{\mathrm{+}}(t)$.

\begin{figure}
\begin{center}
\includegraphics[width=1.1\linewidth]{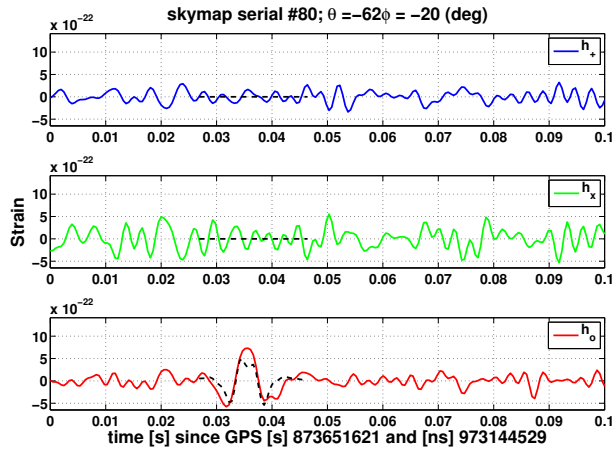}
\caption{Reconstructed polarization modes. The top plot is $h_{\mathrm{+}}$, the middle plot is $h_{\mathrm{\times}}$, the bottom plot is $h_{\mathrm{\circ}}$. The x axis is time with the unit of millisecond since GPS time $873651621.97314453$[s]. The black dashed line in each plot is the injected signal.}
\label{fig:H1L1V1K1_Omega40k}
\end{center}
\end{figure}
Figure~\ref{fig:H1L1V1K1_Omega40k} shows the reconstruction of $h_{\mathrm{+}}$, $h_{\mathrm{\times}}$, $h_{\mathrm{\circ}}$, which is one of segments triggered by the pipeline. In the reconstructed time-series of $h_{\mathrm{\circ}}$, the injected signal is clearly reconstructed. The injected $h_{\circ}$ different from Figure~\ref{fig:scalarGW}. This difference comes from the fact that the low frequency region below $60$~Hz and the high frequency region above $400$~Hz in data are filtered in this simulation. 

In this paper, although our approach does not need gravitational waveforms a priori, we use the SNR calculated in the same way of Eq.~(\ref{eq:def_snr}) in order to evaluate the accuracy of the reconstruction:
\begin{equation}
\mathrm{SNR}=\left(\int^{\infty}_{-\infty}\frac{\sum_A\tilde{\mathfrak{h}}_{A}(f)\tilde{h}_A^*(f)}{S_\mathrm{nA}(f)}\mathrm{d}f\right)^{1/2},
\label{eq:def_snr_R}
\end{equation}
where $\tilde{\mathfrak{h}}_A(f)$ is the reconstructed $h_A$ in Fourier domain, $\tilde{h}_A(f)$ is the injected $h_A$ in Fourier domain, and $S_\mathrm{nA}(f)$ is the power spectrum density of the reconstructed $h_A$.
It should be noted that realistic searches for GWs from supernovae cannot use the waveform models because the models are still not well-established at present. The realistic search may use other indicators which do not use waveform models such as the excess power statistics~\cite{2001PhRvD..63d2003A}. 

The SNR of the reconstructed $h_{\circ}$ which is band-filtered is $14.3$. The SNR of $h_{\circ}$ in H, L, V, K before the reconstruction is $4.9$, $13.4$, $16.4$, $0.54$. $F_\circ$ is $-0.26$ for H, $0.42$ for L, $-0.45$ for V, $-0.01$ for K. The main contribution to the SNR is from V which has best antenna pattern among them.

\begin{figure}
\begin{center}
\includegraphics[width=1\linewidth]{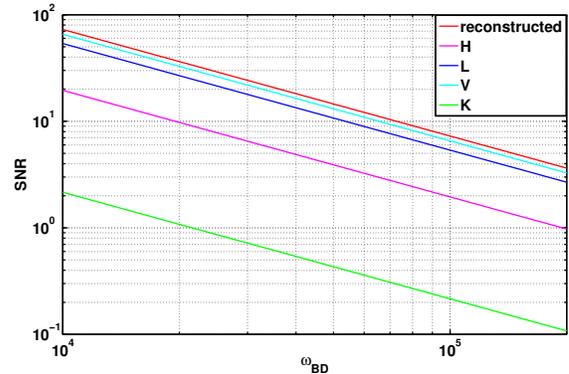}
\caption{The SNR of $F_\circ h_\circ$ in H, L, V, and K, and reconstructed $h_\circ$ as a function of $\omega_{\mathrm{BD}}$.}
\label{fig:snr_plot}
\end{center}
\end{figure}
Figure~\ref{fig:snr_plot} shows the SNR of $h_\circ$ in H (the magenta line), L (the blue line), V (the light blue line), and K (the green line), and reconstructed $h_\circ$ (the red line) as a function of $\omega_{\mathrm{BD}}$. The reconstruction procedure were repeated $28$ times with different data set and the SNRs were calculated at each trial and then they were averaged. Interestingly, the SNR of the reconstructed $h_\circ$ is higher than the others.  
As shown in Eq.~(\ref{eq:ML}), the maximum likelihood method combines all detector data streams that are weighted by their sensitivities in the whitening procedure. Multiple sensitive-detector data streams make contribution to reduce the variance of the reconstructed $h_\circ$ and the SNR of the reconstructed $h_\circ$ rises up.
In regard to constraining $\omega_{\mathrm{BD}}$ by practical GW observations, further studies including the difference of each detector sensitivity and the dependence of SNR on a sky position are needed.

\subsection{Reconstruction of the polarization modes with tensor mode signals}

\begin{figure}
\begin{center}
\includegraphics[width=1.1\linewidth]{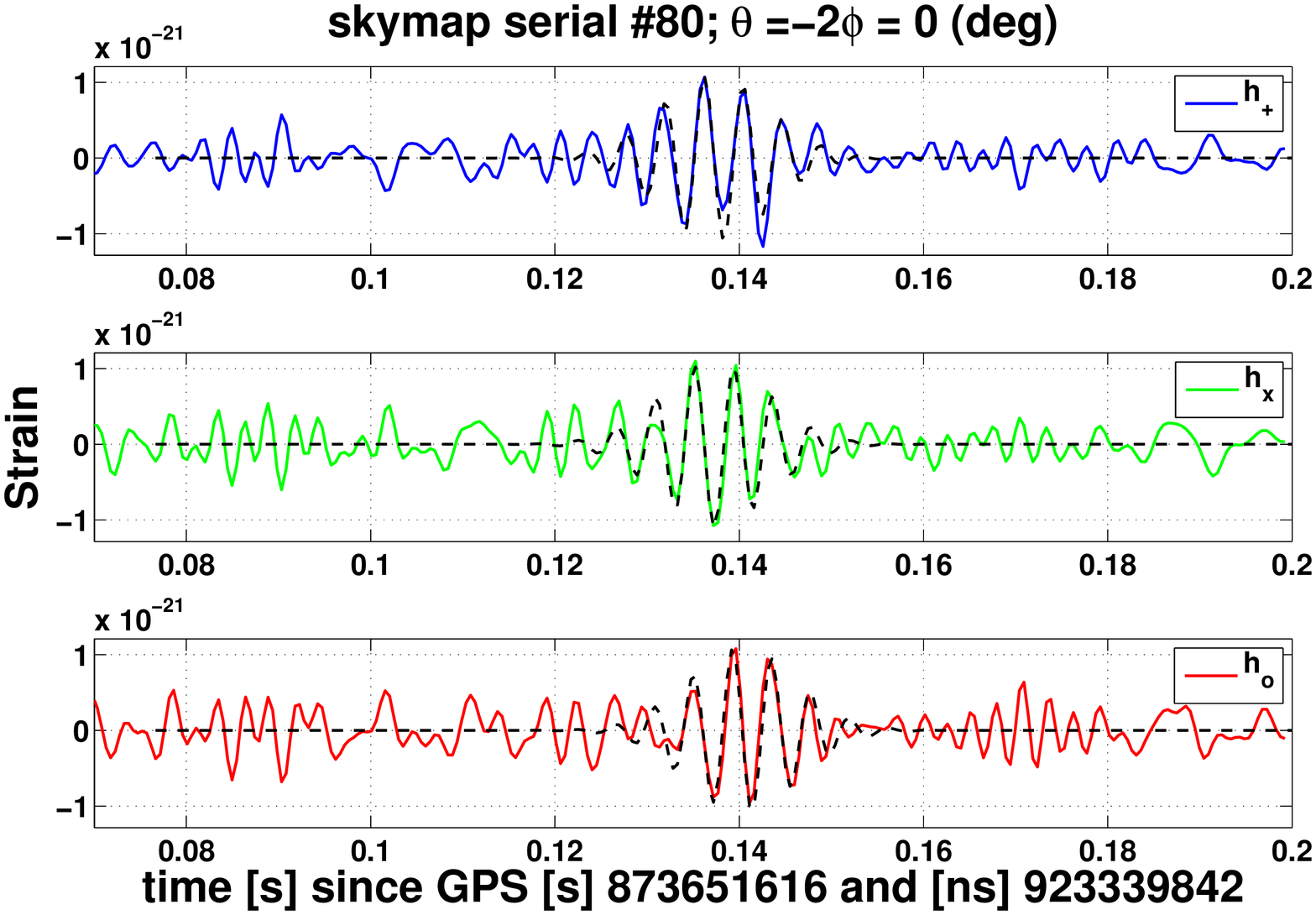}
\caption{
Reconstructed $h_{\mathrm{+}}$, $h_{\mathrm{\times}}$, and $h_{\mathrm{\circ}}$. The black dashed line on each plot is the injected waveform.
\label{fig:code_test}
}
\end{center}
\end{figure}
In previous section we considered the reconstruction of the GW which contains only scalar mode. The algorithm can reconstruct all the modes if they exist. In this section we demonstrate the reconstruction of a GW which contains three modes, $h_+$, $h_\times$, $h_\circ$. Example of the reconstruction of the polarization modes $h_{\mathrm{+}}$, $h_{\mathrm{\times}}$, and $h_{\mathrm{\circ}}$ is shown in Fig.~\ref{fig:code_test}. The detectors H, L, V, K are assumed to have the same sensitivity as shown Fig.~\ref{fig:detector_sensitivity}. The set of detector noise is generated in the same way in section \ref{Simulation}. 
Regarding to injection signals, in this simulation we will not use GW waveforms predicted by the Brans\mbox{-}Dicke theory because spherically symmetric core collapse supernovae do not produce tensor modes, and there is no realistic simulations on asymmetric core collapses in the Brans\mbox{-}Dicke theory. Instead, we will use the sine-Gaussian signals that are usually used for evaluation of GW burst searches.
 The injected signals correspond to a single source located at the right ascension 
(RA) of $16.4~{\rm hours}$ and the declination (DEC) of $0~{\rm degrees}$. We assume that
\begin{eqnarray}
 h_\times(t) &=& A\exp[-(2\pi f_0t)^2/2Q^2]\sin(2\pi f_0t),\nonumber \\
 h_+(t) &=& A\exp[-(2\pi f_0t)^2/2Q^2]\sin(2\pi f_0t+\pi/2),\nonumber \\
 h_\circ(t) &=& A\exp[-(2\pi f_0t)^2/2Q^2] \sin[2\pi f_0(t+0.003)], \nonumber \\
 &&
\end{eqnarray} 
where $Q=9$ is the Q-value and $f_0=235$~Hz is the central frequency. 
The signal strength A is scaled so that the root-sum-square $h_{\mathrm{rss}}=1.38\times 10^{-22}\mathrm{Hz}^{-1/2}$.
The dashed lines in Fig.~\ref{fig:code_test} are the injection signals. We evaluate the signal strength by SNR. Table~\ref{table1} shows the SNR of the reconstructed $h_+$, $h_{\times}$, $h_\circ$, and the SNR before reconstruction of the three polarization modes in H, L, V, K. 

\begin{table}[t]
\begin{center}
\begin{tabular}{|l|c|c|c|c|r|}
	\hline
	            & Reconstructed & H & L & V & K \\ \hline
	$h_+$       & 23.3 & 8.3 & 9.6 & 26.7 & 24.5 \\ 
	$h_\times$  & 20.2 & 15.4 & 18.0 & 15.4 & 10.8 \\ 
	$h_\circ$   & 16.3 & 13.3 & 20.4 & 10.0 & 6.2\\
	\hline
\end{tabular}	
\end{center}
\caption{The SNR of the reconstructed $h_+$, $h_{\times}$, $h_\circ$, and the SNR of the three polarization modes in H, L, V, K. The antenna pattern function $(F_+,F_\times,F_\circ)$ is $(0.25,-0.46,0.39)$ for H, $(0.19,0.36,-0.41)$ for L, $(-0.65,-0.38,-0.24)$ for V, $(0.56,0.25,0.14)$ for K.}
\label{table1}
\end{table}

\begin{figure*}
\begin{center}
\includegraphics[width=.49\linewidth]{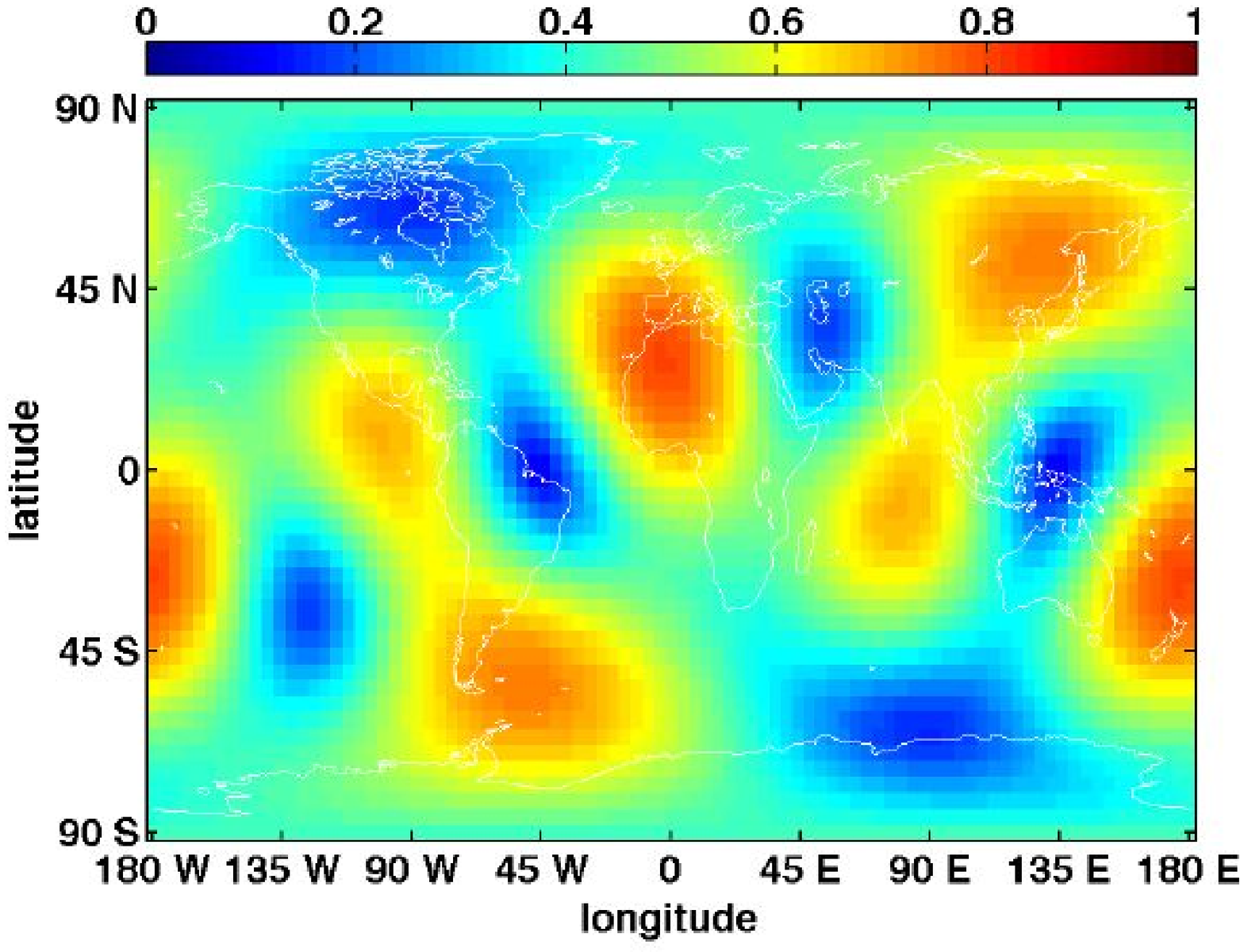}
\includegraphics[width=.49\linewidth]{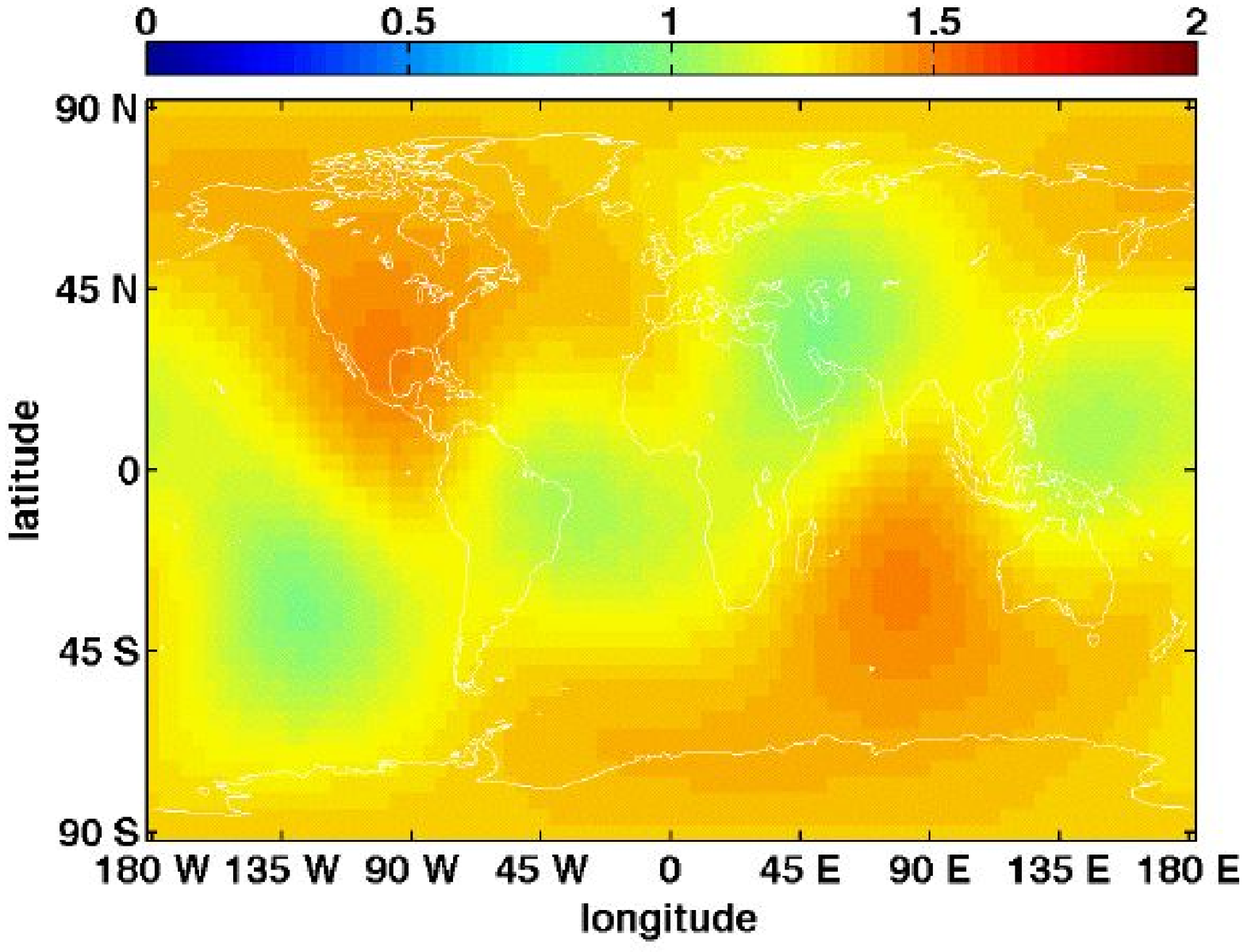}
\includegraphics[width=.49\linewidth]{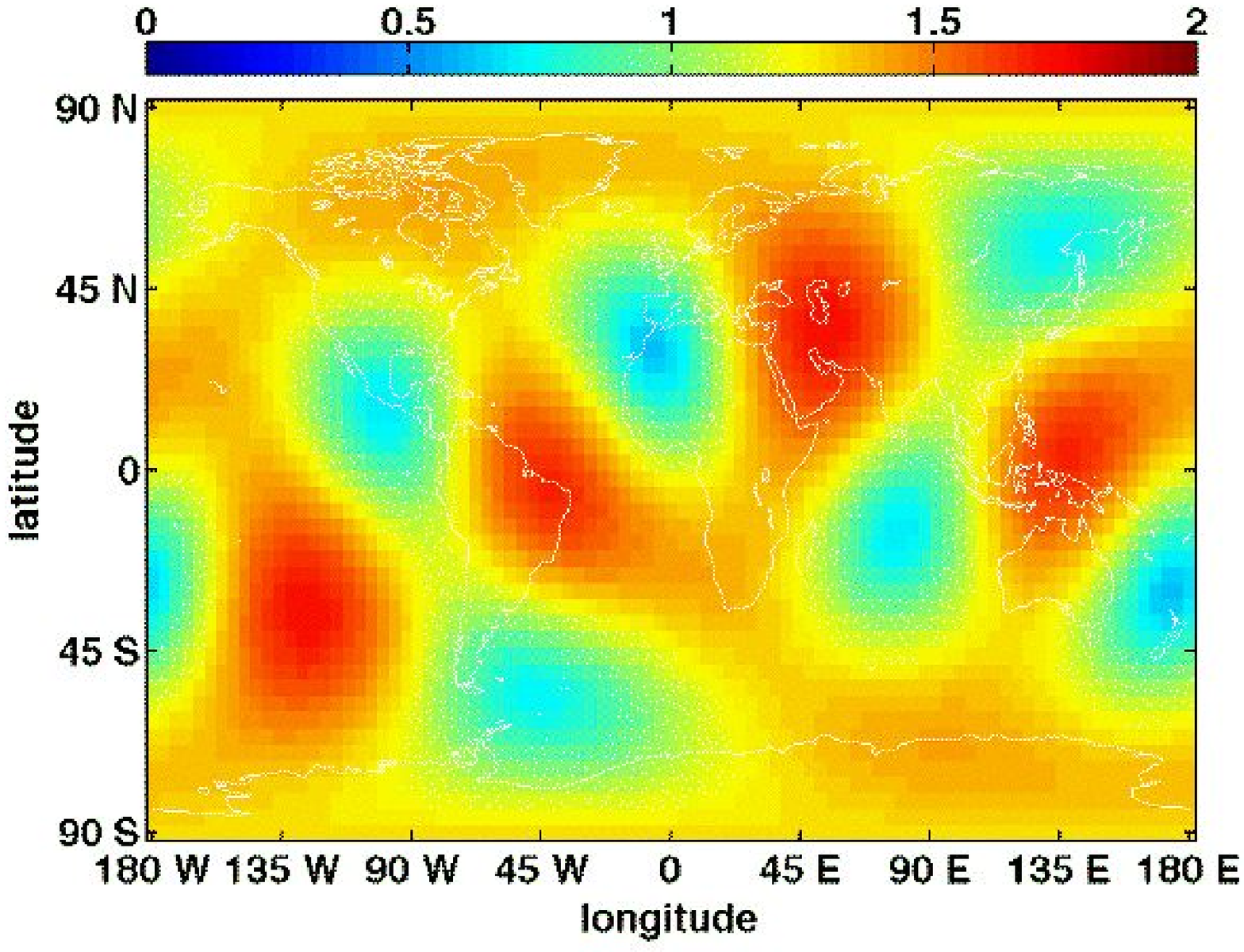}
\caption{The left top plot is the network antenna pattern of H-L-V-K to a scalar gravitational wave.  The $X$-axis is the longitude and the $Y$-axis is the latitude.  The right top plot is the network antenna pattern function to the tensorial modes. The bottom plot is the summed antenna pattern function to the vector modes
\label{fig:network_antenna_pattern}}
\end{center}
\end{figure*}

If we define a mean SNR as 
\begin{equation}
	\overline{\mathrm{SNR}} = \sqrt{\frac{\mathrm{SNR}(h_+)^2+\mathrm{SNR}(h_\times)^2+\mathrm{SNR}(h_\circ)^2}{3}},
	\label{mean_snr}
\end{equation}
$\overline{\mathrm{SNR}}$ is $20.1$ for reconstructed modes, $12.7$ for H, $16.7$ for L, $18.7$ for V, $15.9$ for K. From the Table~\ref{table1}, one can see even if several detectors in the detector network do not have sensitivities to certain polarization modes, sensitive detectors in the detector network can make improvement of the SNR. 
One indication to know the detector-network sensitivity to the polarization modes is the network antenna pattern skymap $\bar{F}_T:=\sqrt(\sum_{I=1}^m F_+^{I2}+F_\times^{I2})$, $\bar{F}_S:=\sqrt(\sum_{I=1}^m F_\circ^{I2})$, $\bar{F}_V:=\sqrt(\sum_{I=1}^m F_x^{I2}+F_y^{I2})$. Figure~\ref{fig:network_antenna_pattern} shows the network antenna pattern skymap of H-L-V-K. Comparing the antenna pattern skymaps of the polarization modes for H (see Figure~\ref{fig:antenna_pattern}), for the antenna pattern skymaps of the tensor modes, the region where the value is above $0.5$ is increased from $72.8$\% to $100$\%, for the vector modes, the region where the value is above $0.5$ is increased from $69.1$\% to $100$\%, and for the scalar mode, the region where the value is above $0.25$ is increased from $31.4$\% to $94.2$\%. 
These shows the benefit of the use of multiple detectors in terms of the improvement of the SNR and the sky coverage. The latter means that angular directions insentive to a certain polarization mode is removed and that the detector network can distinguish all polarizations. It should be noted that the matched filtering method on reconstructed polarization modes is proposed in ~\cite{2008CQGra..25r4021H}. This method benefit from the detector network directly.


\section{Summary}
\label{summary}

This paper develops a method to reconstruct at most five non\mbox{-}Einsteinian polarization modes, in addition to $h_{\mathrm{+}}$ and $h_{\mathrm{\times}}$, of GWs using a network of ground-based interferometric GW detectors. This method is applicable for testing alternative theories of gravity by searching for scalar and vector GWs. Since the method does not rely on any specific model of the  alternative theories and does not need theoretical models of gravitational waveforms, the method can be a model-independent test of alternative theory of gravity.

We overview the algorithm of the method. The detector responses is first whitened by the linear predictor error filter so that the processed data is sample to sample uncorrelated. Since the detector response to a GW is the linear combination of the polarization modes weighted with the antenna pattern functions, the reconstruction of the polarization modes is naturally formulated as an inverse problem. We solved the inverse problem by minimizing residuals by subtracting a reconstructed signal from the detector output data.

We performed simulations of the reconstruction of the polarization modes as demonstrations. We used the simulated gravitational waveform from a spherically symmetric core collapse supernova in Brans-Dicke theory, which predicts only the scalar GW, and showed the scalar GW was well reconstructed. We also demonstrated that all gravitational waveforms were well reconstructed in the presence of two tensor modes in addition to a scalar mode, using sine-Gaussian waveforms. 
The mean SNR defined in Eq.~(\ref{mean_snr}) was calculated for the reconstructed polarization modes and the polarization modes in H, L, V, K before reconstruction. The mean SNR of the reconstructed polarizations were higher than the others, which means that sensitive detectors play an important role to reconstruct the waveforms and gain the SNR.
For further study of the detectability, more realistic simulations, considering the sky position dependence of a source and other waveforms in a specific model of gravity theory, are encouraged.



\begin{acknowledgments}
K.H. is supported by Grant-in-Aid for Young Scientists (B) and the Max-Planck-Society. K.H. would like to thank to Bruce Allen for warm hospitality during his stay in Hannover. K.H.would like to thank to Soumya D. Mohanty for valuable comments and encouragement. A.N. is supported by a grant-in-aid through JSPS. We would like to thank to Atsushi Taruya for fruitful discussion.
\end{acknowledgments}

\bibliography{bransdicke_PRD}

\end{document}